\documentclass[fleqn,usenatbib,letters]{mnras}

\usepackage{newtxtext}
\usepackage{mathptmx}
\usepackage{txfonts}

\usepackage[T1]{fontenc}
\usepackage{ae,aecompl}

\usepackage{graphicx}	

\usepackage{xcolor}

\title[Evolution of Earth-like extended exospheres]{Evolution of Earth-like extended exospheres orbiting solar-like stars}

\author[A. Canet and Ana I. Gómez de Castro]{
Ada Canet$^{1,2,3}$\thanks{E-mail: adacanet@ucm.es}
and Ana I. G\'omez de Castro$^{1,2}$
\\
$^{1}$Joint Center for Ultraviolet Astronomy (JCUVA), Universidad Complutense de Madrid, Madrid, Spain\\
$^{2}$Facultad de Ciencias Matemáticas, U.D. Astronomia y Geodesia, Universidad Complutense de Madrid, Madrid, Spain\\
$^{3}$Facultad de Ciencias Físicas, Departamento de Física de la Tierra y Astrofísica, Universidad Complutense de Madrid, Madrid, Spain
}

\date{Accepted XXX. Received YYY; in original form ZZZ}

\pubyear{2021}

\begin{document}
\label{firstpage}
\pagerange{\pageref{firstpage}--\pageref{lastpage}}
\maketitle

\begin{abstract}
Recent observations of the Earth's exosphere revealed the presence of an extended hydrogenic component that could reach distances beyond 40 planetary radii. Detection of similar extended exospheres around Earth-like exoplanets could reveal crucial facts in terms of habitability. The presence of these rarified hydrogen envelopes is extremely dependent of the planetary environment, dominated by the ionizing radiation and plasma winds coming from the host star. Radiation and fast wind particles ionize the uppermost layers of planetary atmospheres, especially for planets orbiting active, young stars. The survival of the produced ions in the exosphere of such these planets is subject to the action of the magnetized stellar winds, particularly for unmagnetized bodies. In order to address these star-planet interactions, we have carried out numerical 2.5D ideal MHD simulations using the PLUTO code to study the dynamical evolution of tenuous, hydrogen-rich, Earth-like extended exospheres for an unmagnetized planet, at different stellar evolutionary stages: from a very young, solar-like star of 0.1 Gyr to a 5.0 Gyr star. For each star-planet configuration, we show that the morphology of extended Earth-like hydrogen exospheres is strongly dependent of the incident stellar winds and the produced ions present in these gaseous envelopes, showing that the ionized component of Earth-like exospheres is quickly swept by the stellar winds of young stars, leading to large bow shock formation for later stellar ages.
\end{abstract}

\begin{keywords}
planets and satellites: atmospheres -- MHD -- methods: numerical
\end{keywords}



\section{Introduction}
The Earth's exosphere is mainly constituted by atomic hydrogen (HI) atoms, gravitationally bound to our planet describing ballistic, satellite and escaping trajectories \citep{1963P&SS...11..901C} in a quasi-collisionless environment. Terrestrial exospheric HI atoms result from the photodissociation of water, methane and/or molecular hydrogen driven by the ultraviolet (UV) radiation at lower atmospheric layers, in turn diffused into the upper atmosphere \citep{1994GeoRL..21.2563D,Brasseur1996}. In this sense, the detection of a hydrogen envelope around a distant planet could be used as an indirect measurement of the presence of greenhouse gases in the lower atmosphere, or even the presence of water oceans \citep{2007Icar..191..453S, 2017MNRAS.464.3728B}.

Detection of H-rich envelopes around Earth-like exoplanets demands for UV observations, as the most sensitive tracer to detect neutral atmospheric hydrogen is the imprint left by the absorbing gas in the Ly-$\alpha$ line during the transit of the planet. Close-in gaseous giant exoplanets and Earth-like planets provided with massive H-rich, nebula-based protoatmospheres \citep{1978PThPh..60..699M,1979E&PSL..43...22H} have been widely modeled, as they give the best chances to detect powerful hydrodynamic escape driven by the extreme UV stellar flux \citep{2013AsBio..13.1011E, 2013AsBio..13.1030K, 2014MNRAS.439.3225L,2018MNRAS.479.3115V,2019ApJ...873...89M, 2020MNRAS.498L..53C}. However, the detection of an Earth-like hydrogenic exosphere around a distant exoplanet would be challenging due to their low-density profiles, in comparison to the high-density, massive nebula-based hydrogen protoatmospheres around Earth-like planets or gas giant escaping atmospheres.

Recent observations of the Earth's outer atmosphere have improve the prospects of detection. The observations carried out by the Geocorona Photometer (GEO) far ultraviolet (FUV) spectral imaging instrument on board the Imager for Magnetopause to Aurora Global Exploration (IMAGE) \citep{2000SSRv...91...51F} showed the presence of an extended exospheric component, mainly populated by satellite particles describing elliptic trajectories \citep{2000SSRv...91..287M}, further confirmed by the Two Wide-angle Imaging Neutral Atoms Spectrometer (TWINS) mission \citep{2009SSRv..142..157M}. Following these results, the Lyman Alpha Imaging Camera (LAICA) on board the Japanese micro-spacecraft PROCYON\footnote{Proximate Object Close Flyby with Optical Navigation.}  
\citep{2017GeoRL..4411706K} revealed that the neutral hydrogen exosphere around the Earth reaches more than 40 radii, a distance that could be even larger according to the Solar Wind Anisotropies/Solar and Heliospheric Observatory (SWAN/SOHO) instrument measurements \citep{2019JGRA..124..861B}. Extended exospheres (hereafter EEs) are also present in Mars and Venus \citep{1996RvGeo..34..483S}, showing that these gaseous envelopes are a common feature in the rocky planets of the Solar System. Hence, it is reasonable to expect them in other Earth-like exoplanets orbiting cool main-sequence stars like the Sun. Recently, \citet{2018ExA....45..147C} showed the feasibility of detecting Earth-like exospheres with moderate size (4–8 m primary mirror) space observatories through transit spectroscopy in the Ly-$\alpha$ line. This result was later confirmed by \citet{2019A&A...623A.131K} and \citet{2019A&A...622A..46D}. 

The abundance of Earth-like exospheres in other planetary systems depends strongly on their survival under the joint effect of the X-ray and EUV stellar radiation, and the particle flow coming from the host star, i.e. stellar winds. The characteristic velocity, density, temperature and magnetic fields of such these winds strongly depend on the stellar activity, and hence, exoplanets orbiting more active stars will be hit by stronger stellar winds. For non-magnetized planets, stellar winds and photoionizing radiation directly interact with the planetary exosphere, ionizing the neutral species. The magnetic field carried by the stellar wind may divert efficiently the ionized particles from the exosphere preventing the survival of large exospheres. The interaction between the stellar radiation and/or plasma winds with the atmospheres of unmagnetized or weakly magnetized exoplanets, and the consequent ion losses, has been widely modeled in the case large, nebula-based H-rich exospheres \citep{2005ApJS..157..396E,2013AsBio..13.1011E,2013AsBio..13.1030K,2014A&A...562A.116K,2015ApJ...808..173T,2016ApJ...832..173S,2017ApJ...847..126K}, and thicker CO2-rich atmospheres like those present in Mars and Venus \citep{2009AsBio...9...55T, 2013JGRA..118..321M,2014JGRA..119.1272M,2007GeoRL..34.8201M,2015ApJ...806...41C, 2017ApJ...836L...3A, 2018GeoRL..45.9336S, 2019MNRAS.488.2108E}. However, the above described Earth's observations are not included in these models often used to evaluate the detectability of the atmospheres of Earth-like exoplanets.

In this work, we show the results of numerical, single-fluid MHD simulations for the study of the dynamical evolution of the ionized hydrogen present in the EE of an unmagnetized Earth-like planet ($R_p=$1.0$R_{\oplus}$ and $M_p =1.0M_{\oplus}$, where $R_{\oplus}=6.37\times 10^6$m and $M_{\oplus}=5.97\times 10^{24}$kg), considering different initial exospheric densities, and exposed to the action of the stellar wind coming from a solar-like star (1.0$M_{\odot}$, 1.0$R_{\odot}$) at 1.0 au. To properly describe the stellar winds at different stellar activity levels, we obtained the characteristic wind parameters corresponding to different stellar ages (from a very young 0.1 Gyr star up to 5.0 Gyr).
In Section 2 we describe the solution of the stellar winds in order to obtain the characteristic wind parameters for different stellar ages. In Section 3 we describe the numerical MHD code used in this work, including the initial and boundary conditions of our numerical setup. Finally, we report the results of the numerical simulations, showing the characteristic swept-up time for each star-planet configuration and the steady state for different stellar ages and initial exospheric densities.  We show that the dynamic interaction of the stellar wind with the ionized exospheric particles leads to the formation of extended bow shocks at later stellar ages, removing all the ionized material from planets orbiting young solar-like stars.

\section{Stellar Winds}\label{sec:winds}
The stellar activity is reflected in the characteristic velocity, density, temperature and magnetic field of the corresponding stellar winds. As planetary upper atmospheres, especially in the case of non-magnetized planets, are exposed to the action of such these winds, their evolution must be studied according those parameters.

In order to obtain the wind characteristic parameters for a solar-like (1.0$M_{\odot}$, 1.0$R_{\odot}$) star at 1.0 au, we solved the analytical isothermal model of \cite{1967ApJ...148..217W} (hereafter WD) for the solar wind, following the method described in \citet{2005A&A...434.1191P}, also used in \citet{2017A&A...598A..24J} for validating their numerical model, where the rotation rate of the star $\Omega_{\star}$, the radial component of the magnetic field at the stellar surface $B_{r0,\star}$, the mass loss rate $\dot{M}_{\star}$ and the base-wind temperature of the star $T_{wind}$ are used as input parameters. As the stellar activity could be related to the stellar age, the solutions have been computed to cover a broad range of ages, from 0.1 Gyr (very active star), up to 5.0 Gyr (Sun-like stellar activity), by using the stellar age-dependent scale relations into the input parameters: for $\Omega_{\star}$ and $B_{r0,\star}$ we used the relations given in \citet{2014MNRAS.441.2361V}; for the estimation of $T_{wind}$ and $\dot{M}_{\star}$, we use those derived by the MHD wind simulations carried out in \citet{2018MNRAS.476.2465O}, where a change in the trend of such parameters with stellar age is predicted for stars older than 2.0 Gyr.

However, the isothermal assumption considered in the WD model leads to excessive values of the temperature at distances far from the star's surface. It is usual in stellar wind models to use the polytropic approximation. Based on empirical data from solar wind observations \citep{2011ApJ...727L..32V}, we have opted to evaluate the wind temperature at 1.0 au using the polytropic approach with index $\gamma=1.1$.

Values for the obtained characteristic parameters of the stellar wind at 1.0 au, such as numerical density $n$, radial velocity $v$, magnetic field $|B|$, mass-loss rate $\dot{M}$ and electronic temperature $T_e$ are given in Table \ref{tab:sw}. For all considered stellar ages, winds at 1.0 au are in the super-Alfv\'enic regime, with sound Mach numbers $M_s=|v|/c_s$ between 6.38, 
in the case of the eldest star and 7.91 for the youngest. The Alfv\'enic Mach number ($M_A=|v|/v_A$, with $v_A$ the Alfv\'einc speed) and the magnetosonic Mach number ($M_f=|v|/\sqrt{v_A^2+c_s^2}$) take values between 74.0 and 15.7, and between 7.0 and 5.2 respectively, the lowest values corresponding to a 5.0 Gyr star, and the highest to a 0.1 Gyr star.
\begin{table}
\caption{\textbf{Stellar wind parameters at 1.0 au for different stellar ages.} }
\begin{tabular}{c|ccccc}
\hline
t  & n  & v  & |B|  & $\dot{M}$  & T$_e$ \\
(Gyr) & (cm$^{-3}$) & (km/s) & (G) & (M$_{\odot}$/yr) & (MK) \\\hline
0.1     & 2104        & 1168     & $1.1x10^{-3}$ & $9x10^{-12}$            & 1.2      \\
1.0     & 550         & 894      & $1.2x10^{-4}$ & $1.8x10^{-12}$           & 0.8      \\
3.0     & 61          & 498      & $6.0x10^{-5}$ & $1.1x10^{-13}$           & 0.3      \\
5.0     & 10          & 407      & $4.2x10^{-5}$ & $1.5x10^{-14} $          & 0.2      \\ \hline
\label{tab:sw}
\end{tabular}
\end{table}

\section{Numerical simulation of wind-exosphere interaction}
The interaction between the stellar wind coming from a cool, solar-like star at 1.0 au and a non-magnetized EE has been modeled for stellar ages of 0.1, 1.0, 3.0 and 5.0 Gyr and varying the initial exospheric ion density, resulting in a total number of 16 numerical simulations to study this interaction. The evolution of the EEs is studied by comparing the steady state obtained for each of these configurations. The simulations have been carried out using the open source code 
PLUTO \citep{2007jena.confR..96M}, a modular Godunov-type code, designed to solve systems of conservation laws in astrophysical fluid dynamics. For these simulations, the ideal MHD module has been used to solve the ideal, single-fluid time-dependent MHD equations:
\begin{equation}
    \frac{\partial \rho}{\partial t}+{\nabla} \cdot (\rho \textbf{v})=0 
    \label{masa}
\end{equation}

\begin{equation}
    \frac{\partial(\rho\mathbf{v})}{\partial t} + \nabla (\rho \mathbf{v}\mathbf{v}-\mathbf{B}\mathbf{B})+\nabla p_t = \rho \mathbf{g}
\end{equation}

\begin{equation}
    \frac{\partial E}{\partial t} + \nabla ((E+p_t)\mathbf{v}-\mathbf{B}(\mathbf{v}\cdot\mathbf{B})) = \rho \mathbf{v}\mathbf{g}
\end{equation}
    
\begin{equation}
    \frac{\partial \mathbf{B}}{\partial t} + \nabla (\mathbf{v}\mathbf{B}-\mathbf{B}\mathbf{v}) = 0
\end{equation}

where $\rho=\mu n_{tot} m_p$ is the mass density, with $\mu$ the fraction of ionization (here we will consider the stellar winds such a fully ionized gas, so we will take $\mu$=0.5), $m_p$ the proton mass and $n_{tot}$ the total number of particles; $\mathbf{v},\mathbf{B}$ and $\mathbf{g}$ are the velocity, magnetic field and acceleration vector respectively. The total pressure of the system is $p_t=p+\mathbf{B}^2/2$, including thermal and magnetic pressures and $E$ is the energy of the system, defined as $E=\rho \epsilon + \frac{\rho \mathbf{v}^2}{2}+\frac{\mathbf{B}^2}{2}$, where $\epsilon$ is the internal energy.\\
For the simulation of the interaction between the stellar wind and planetary exosphere under the MHD approximation, we consider the stellar wind as a fully ionized fluid consisting mostly of protons and electrons, with an embedded magnetic field. Fluid models are suitable for describing the large-scale attributes of the interaction between magnetized plasmas and the ionospheric component of a planetary atmosphere, including the produced wind disturbances such as shocks and the formation of magnetic barriers around the obstacle (planet), and that's the aim if this work. A more accurate treatment of the problem would require to solve the plasma physics equations, including the statistical approach to deal with charged particles and neutrals, together with the complex fluid mechanics involving gravity and magnetic fields. For this reason, the theoretical works published till now on the star-planet interaction only provide partial studies of the relevant phenomena, e.g., neglect the impact of the stellar wind magnetization and solve only the hydrodynamical interaction (e.g. \citet{2016ApJ...832..173S}, \citet{2017ApJ...847..126K}), consider only the effect of stellar radiation on atmospheric heating and expansion neglecting collisional ionization by the stellar wind particles (\citet{2013AsBio..13.1030K}), consider the effect of XUV radiation in driving planetary winds neglecting the impact of the stellar wind (\citet{2013MNRAS.430.1247L}, \citet{2015Icar..250..357C}), not considering a the presence of an expanded planetary exosphere (e.g. \citet{2014ApJ...795...86S}), and so on.

MHD equations are used in conjunction with an ideal gas equation of state. The conservative form of these equations is integrated using a Harten, Lax, Van Leer approximate Riemann solver (hll), using a linear reconstruction associated with a 
Van Leer harmonic mean limiter. Second order in time is achieved using a Runge-Kutta scheme, while the divergence of the magnetic field is ensured by Powell's Eight-Waves formulation \citep{powell1994approximate}, where magnetic fields retain a cell average representation.

We use a system of cartesian coordinates, adopting a 2.5D formalism. The planet is defined as an internal boundary of 1.0$R_{p}$, located at the center of the cartesian grid. Defining $x$ as the horizontal dimension, and $y$ as the vertical dimension, the computational grid has an extension,
\begin{equation}
    -x_0\leq x \leq x_0 \qquad -y_0\leq y\leq y_0
\end{equation}
\noindent
where $x_0=y_0=50R_{p}$. We used 512 grid point in each dimension, resulting in a resolution of the computational grid of $\sim 0.2 R_p \times 0.2R_p$.
The initial conditions of our simulation are defined by the density profile of the planetary exosphere, described in detail in the next subsection. The stellar wind velocity follows the $x$ axis (horizontal) direction, traveling from the left boundary side of the cartesian grid, where the stellar wind parameters (velocity, density, pressure and magnetic field), obtained from the WD analytical solutions for each stellar age, are set as boundary conditions. We follow a simple magnetic field orientation approach, considering the interplanetary magnetic field oriented in the positive direction of the $y$ axis (vertical direction), i.e. perpendicular to the speed of stellar wind propagation, and perpendicular to the orbital plane,at the outer boundary of the simulation grid.  Outflow conditions are applied to the rest of the boundaries of the computational domain, i.e.:
\begin{equation}
    \frac{\partial q}{\partial n} = 0, \qquad \frac{\partial \mathbf{v}}{\partial n} = 0, \qquad \frac{\partial \mathbf{B}}{\partial n} = 0
\end{equation}
where \textit{q} is a scalar quantity, \textbf{v} and \textbf{B} are the velocity and magnetic field respectively, and \textit{n} is the coordinate direction orthogonal to the boundary plane.

\subsection{Initial conditions: The planetary exosphere}
To describe the distribution of H ions in the planetary exosphere at the initial timestep of the simulation, we followed a semi-empirical approach based on the known distribution of particles in the Earth's EE as, unfortunately, the properties of the EE cannot be implemented consistently from current planetary wind models. For instance, the exobase level is predicted to be at 7.5$R_{\oplus}$ for a high-density, nebula-based exosphere irradiated by 1XUV and at 9.5$R_{\oplus}$ for 5XUV \citep{2013AsBio..13.1011E}, which are very high values compared with the 1.01$R_{\oplus}$
Earth's exobase level. 

In the initial conditions of the simulation, the EE is assumed spherically symmetric and with a density distribution $n$ based on the \citet{2003JGRA..108.1300O} (hereafter O03) empirical distribution,

\begin{equation}
n(r)=\kappa \Bigg[10^4\;exp\left(\frac{-r}{1.02R_{p}}\right)\\+70\;exp\left(\frac{-r}{8.2R_{p}}\right)\Bigg] {\rm cm}^{-3}, 
\label{eq:exosfera} 
\end{equation}

where $r$ is the radial distance to the center of the planet. This distribution shows two distinct components: the first term of equation [\ref{eq:exosfera}] describes the densest part of the exosphere, located close to the base, mainly populated by ballistic and escaping particles (i.e. a planetary wind), with density $10^4$ cm$^{-3}$; the second term corresponds to an extended exospheric population of satellite particles, with much lower density (70 cm$^{-3}$) that peaks at $\sim8R_{\oplus}$.

\begin{figure}
    \centering
    \includegraphics[width=8cm,height=8cm]{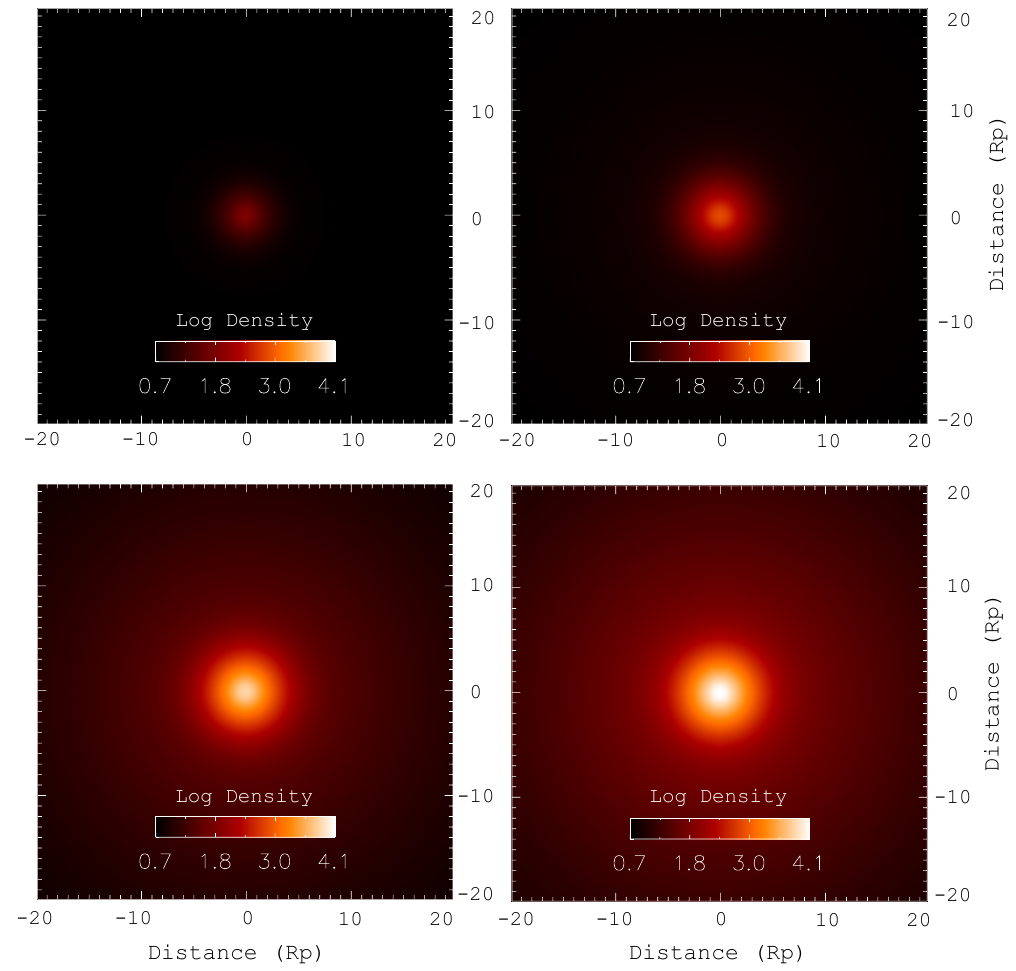}
    \caption{Initial exospheric density distribution for $\kappa=$0.01, 0.1, 1, and 2.0}
    \label{fig:init_exo}
\end{figure}

The $\kappa$ parameter is introduced to scale the ionized particles present in the planetary exosphere. The presence of ionized material in the exosphere of a planet is conditioned by the balance between ion production and losses. Ionization of neutral hydrogen atoms in the uppermost atmospheric layers is a complex process involving a large number of physical mechanisms, including photoionization by stellar Ly-$\alpha$ photons, collisions with the stellar wind electrons, charge-exchange reactions between neutral exospheric particles and fast stellar wind protons, and radiative recombinations. 
In order to simplify the number of ions produced in the planetary exosphere attending to the mentioned processes, we set different initial values for the produced ions through the scale factor $\kappa$ in equation [\ref{eq:exosfera}], introduced to evaluate the impact of the EEs ion density. The $\kappa$ parameter is expected to scale with the planet's atmospheric neutrals, as well as with the density of the stellar wind, the stellar radiation, and hence with the stellar age. The values adopted for $\kappa$ in the simulations are: 0.01, 0.1, 1.0, and 2.0 (see Fig.\ref{fig:init_exo}). Neither the reported variations in the exospheric density distribution with the solar cycle are
considered \citep{2009SSRv..142..157M}.

To provide a soft transition in the border of the internal boundary described by the planetary obstacle, we adapt the density profile inside the planet by a gaussian function of mean $\mu$=0 and variance $\sigma^2$=1.

In the initial conditions of the simulations, the thermal exospheric pressure is defined as
\begin{equation}
p_{exo}=\frac{c_{s,exo}^2 \;\rho(r)}{\gamma},
\end{equation}
where the sound speed $c_s$ is calculated according to the temperature at the Earth's exobase, that we set in 1500K.  

\begin{figure}
\centering
 \includegraphics[width=6cm,height=4cm]{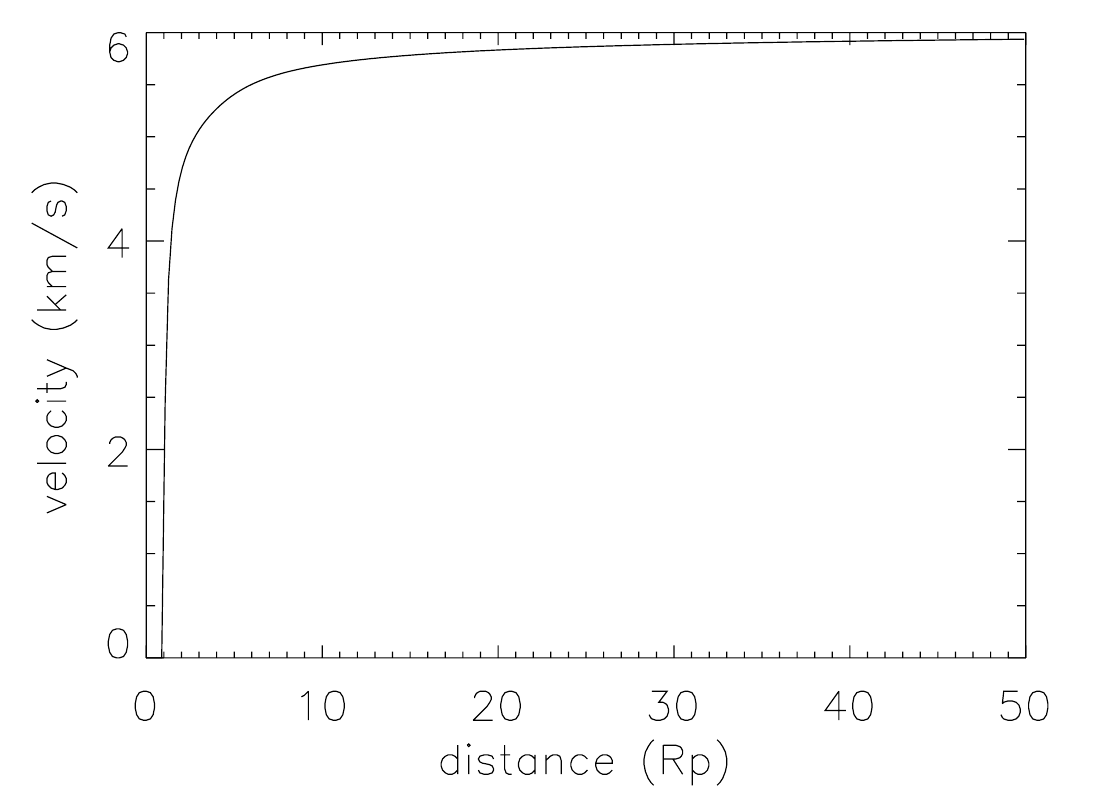}
    \caption{Velocity profile of the planetary outflow considered in the initial conditions of the simulation.}
    \label{fig:wind}
\end{figure}

A planetary wind flow traveling radially from the planetary surface was included in the initial conditions of the simulation. As can be seen in the velocity profile showed in Figure \ref{fig:wind}, the velocity takes values $\sim$4 km/s (corresponding to the sound speed for a temperature of 1500K in the exobase) near the planetary surface, which is consistent with the density and pressure profile considered. The presence of this planetary wind do not affect the final steady state of the simulation, as the dynamic of the system is completely dominated by the stellar wind, due to the high velocity and density values of this plasma. For this reason, we considered the same value for the velocity of the planetary wind present in the simulation domain for all values of $\kappa$ and for all stellar ages.

The magnetic field is set to zero everywhere in the initial conditions, and this value will be constant in the inside domain of the planetary boundary during the simulation.

\section{Results}

\begin{figure*}
\centering
 \includegraphics[width=17cm,height=10cm]{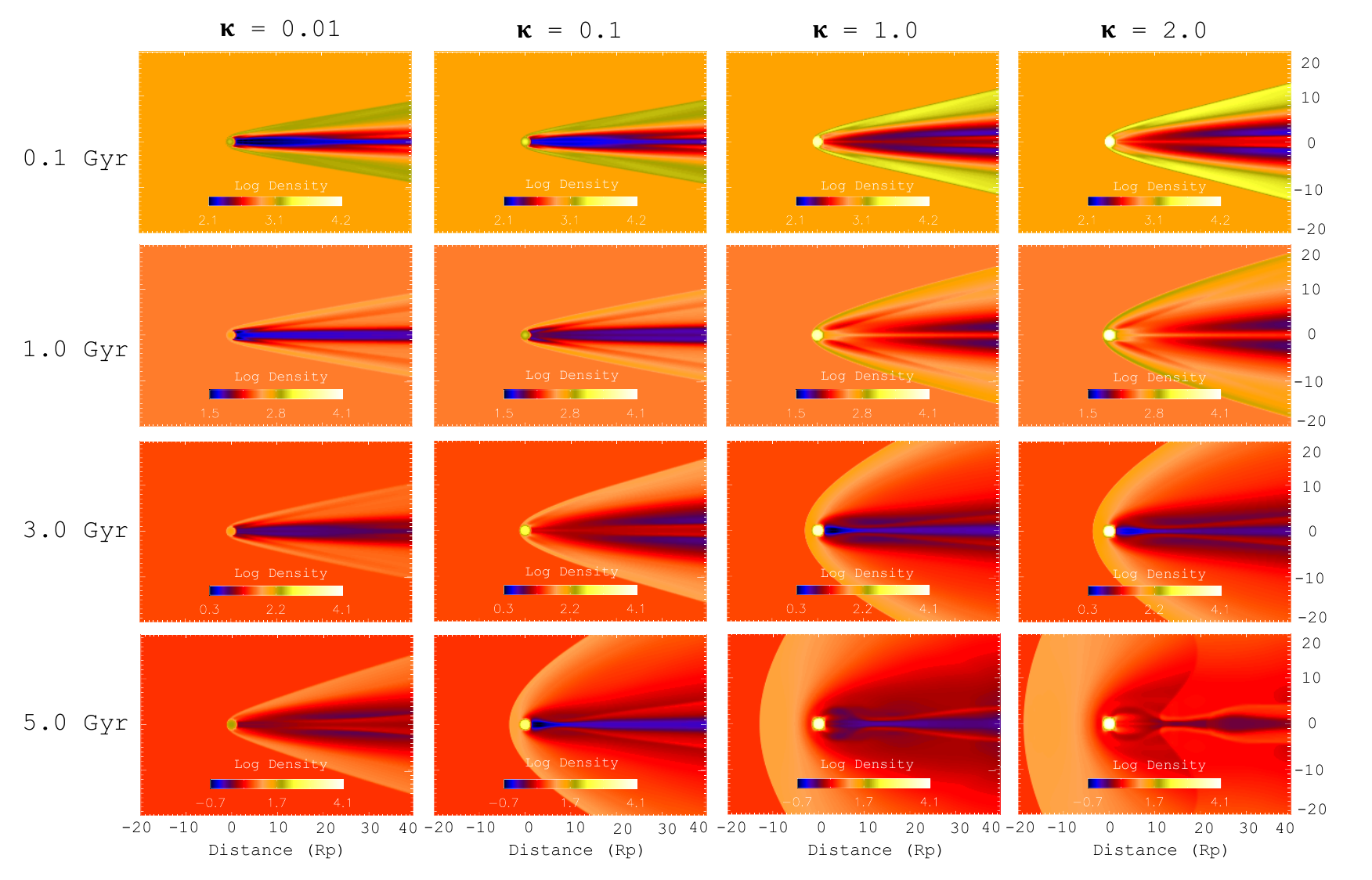}
    \caption{Density distribution in logarithmic scale for each planet-star configuration. Columns from left to right correspond to different exospheric densities: $\kappa=0.01$, $\kappa=0.1$, $\kappa=1.0$ and $\kappa=2.0$ respectively. Rows from top to bottom corresponds to stellar ages of 0.1, 1.0, 3.0 and 5.0 Gyr. Note that different maximum and minimum values for the logarithmic scale change from one stellar age to another.}
    \label{fig:new_mosaic}
\end{figure*}

The final density distribution of exospheric H ions at the steady state of the simulation for each considered star-planet configuration is shown in Figure \ref{fig:new_mosaic}. The morphology of the exosphere is clearly dependent of the parameters of the stellar winds, where the location of the formed bow shock is one of the most remarkable features.

Bow shock formation is a direct consequence of the interaction between the supermagnetosonic flow (stellar wind) and the obstacle (planet). As described in Section \ref{sec:winds}, the stellar winds over the stellar age range considered in this work are in the magnetosonic regime, hence bow shock formation is expected in all studied configurations. The position of the bow shock is determined by the pressure balance on either side of the shock. The ram pressure, defined as $P_{ram}=\rho v^2$, dominates over the stellar wind due to the high densities and velocities of this plasma, being these parameters higher at early stellar ages (0.1 Gyr and 1.0 Gyr), as described in Section \ref{sec:winds}. Ram pressure values range from $4.8\times10^{-5} dyn/cm^2$ for a 0.1 Gyr star, to $2.8\times10^{-8} dyn/cm^2$ for a 5.0 Gyr star. At the initial state of the simulation (t=0), ram pressure of the implemented planetary wind has a value of $2.6\times 10^{-11} dyn/cm^2$ at 10 planetary radii for $\kappa=2$. For the young stars evaluated in our simulations, the stellar wind ram pressure at the initial state of the simulation is at least 6 orders of magnitude higher than the planetary wind ram pressure. In that case, the stellar wind rapidly pushes the planetary atmosphere down to very low atmospheric altitudes, making the bow shock not perceptible in our simulations, as shown in the two first rows in Figure \ref{fig:new_mosaic}; the presence of massive exospheres do not make any significant impact on the interaction between the wind and the planet for early stellar ages.

\begin{figure*}
\centering
\includegraphics[width=16cm,height=11cm]{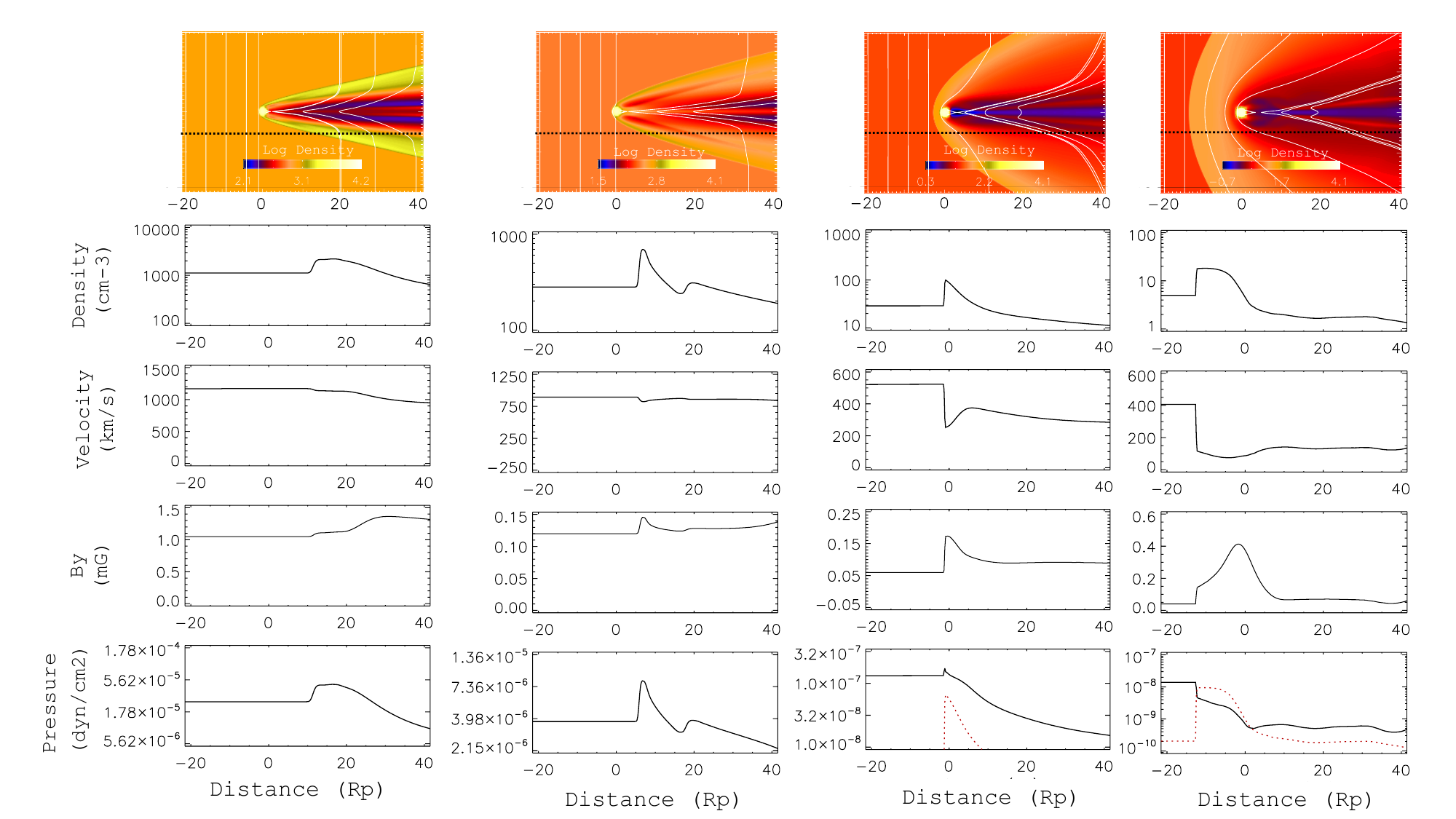}
    \caption{First row: Density distribution in logarithmic scale. White solid lines corresponds to the magnetic field lines. Note that different maximum and minimum values for the logarithmic scale change from one stellar age to another. Second, third, fourth and fifth rows show the density, velocity, tangential component of the magnetic field (By) and pressure profiles respectively, measured at y$=5R_p$ from the center of the planet, as marked in the first row by a dashed black line, for the case of $\kappa=1.0$. Columns from left to right correspond to the cases of 0.1, 1.0, 3.0 and 5.0 Gyr stars, respectively.}
    \label{fig:profiles}
\end{figure*}

For late stellar ages, bow shock extension increases, finding the most extended bow shocks for high-ion density exospheres. As we can observe from density, velocity and magnetic field profiles through the shock (see Fig.\ref{fig:profiles}), the shock produced in all star-planet configurations is a fast shock. A fast shock is characterized by: (1) an increase in density through shock, (2) an increase in the tangential component of the interplanetary magnetic field and (3) a decrease in the normal component of velocity. For the more aged stars of our sample, the difference between the ram pressures of the stellar and the planetary wind is not as marked as in the case of the younger stars. The increase of the density through the shock (and therefore the thermal pressure) is appreciable in these cases, producing more extended bow shocks, as showed in the lowest panels in Figure 4. The presence of massive ionized exospheres also contributes to the net pressure at the planet-side of the shock, producing the most extended bow shocks for more massive exospheres, reaching $\sim$18 planetary radii for ion density characterized by $\kappa=2.0$ orbiting a star of 5.0 Gyr. For such these shocks, the relative density $\rho_1/\rho_0$ between the upstream $\rho_0$ and the downstream density $\rho_1$ is $\rho_1/\rho_0 \sim$3.5.

\begin{figure}
    \centering
    \includegraphics[width=7cm,height=6cm]{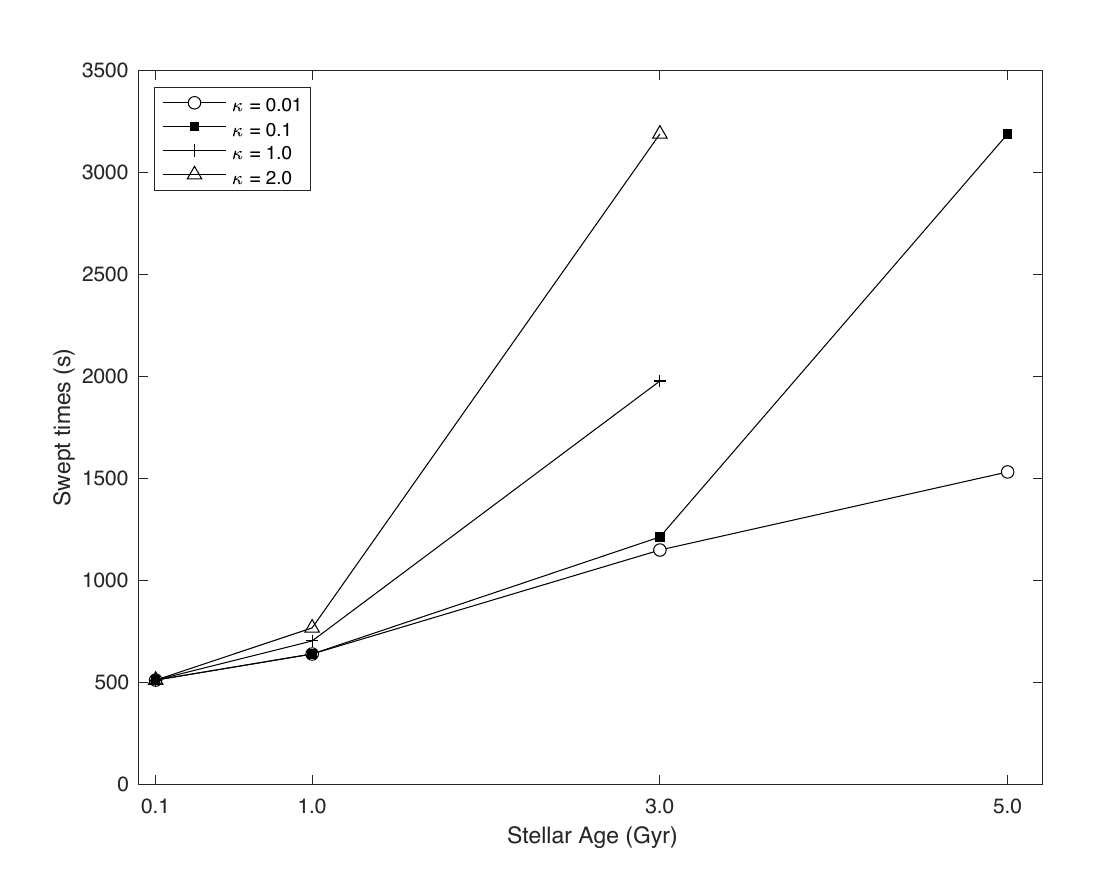}
    \caption{Characteristic exospheric swept-up times $t_{swept}$ measured at 10 $R_p$, as a function of the stellar age, for initial exospheric densities of $\kappa=0.01$, $\kappa=0.1$, $\kappa=1.0$ and $\kappa=2.0$. Swept-up times for 5.0 Gyr and $\kappa$=1.0, 2.0 are not shown in the figure, as the density measured at 10 Rp does not reach interplanetary density values due to the formation of a bow shock.}
    \label{fig:times}
\end{figure}

The exosphere's ionized component of planets orbiting young stars is quickly swept by the action of the stellar winds. The time to remove the ionize material has been measured by defining several sample regions in the computational domain located at 2, 5, 10, 20 and 40 $R_p$, and measured
the time $t_{swept}$ for the exospheric density to drop to interplanetary density values at those locations from the first contact with the stellar wind shock. These times increase with stellar age and initial ion distribution in the planetary exosphere (see Fig. \ref{fig:times}). The ion loss of planetary exospheres is therefore strongly dependent on the ionization state of those gaseous envelopes and the stellar wind conditions.


\section{Discussion and Summary}
\begin{figure}
\centering
 \includegraphics[width=8cm,height=7cm]{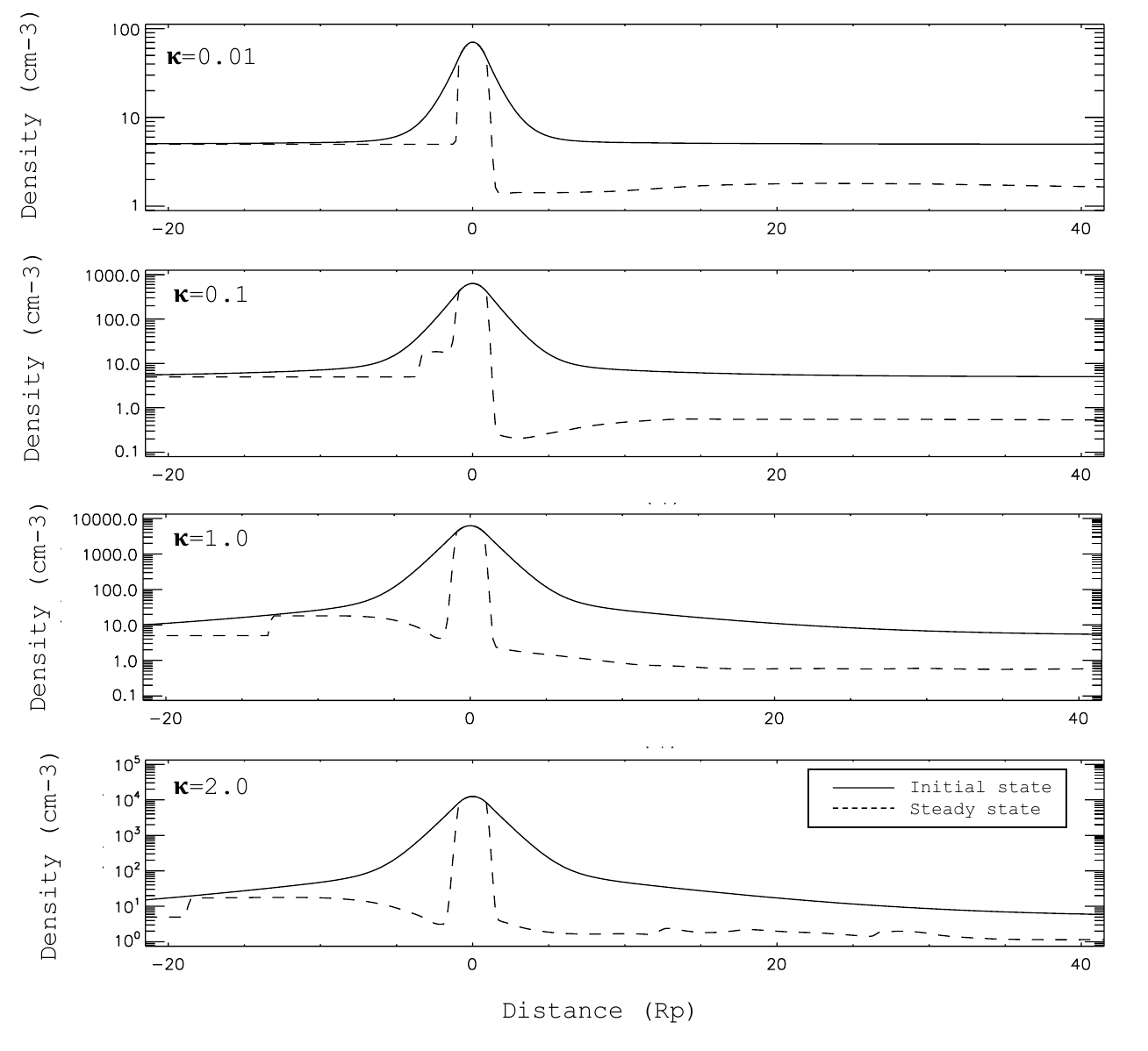}
    \caption{Density distribution profiles in the initial conditions (solid line) and the steady state (dashed line) of the simulation, for a fixed stellar age of 5.0 Gyr and different exospheric densities: $\kappa=0.01$, $\kappa=0.1$, $\kappa=1.0$ and $\kappa=2.0$.}
    \label{fig:inifin}
\end{figure}
The numerical simulations shown above indicate that the stellar wind plays a fundamental role in the spatial configuration of the exospheric ions present in the planetary exosphere. Large bow shocks (up to several planetary radii) are formed for Earth-like planets orbiting solar-like stars of 3.0 and 5.0 Gyr, being the extension of these bow shocks larger for more massive exospheres. Hence, the distribution of ionized hydrogen around these planets is strongly conditioned not only by the strength of the stellar winds, but also the produced ions, or the ionization state of the planetary exosphere.\\
The planetary exosphere is quickly swept against the action of powerful winds coming from young solar-like stars, independently of the ions produced in the planet's exosphere. In that scenario, the density at close distances of the planet is that of the stellar wind as shown in Figure  \ref{fig:inifin}. However, in the case of more evolved stars, the formation of large bow shocks in the case of more massive exospheres leads to an increase of the exospheric density. Hence, even though the interaction with the stellar winds produces a reconfiguration in the shape of the ionized planetary hydrogen exospheres, the values of the density at the places where the shock is formed remain almost constant (see also Figure \ref{fig:inifin}).

Detection of H-rich, Earth-like exospheres has been tested for Earth-like planets orbiting small, nearby cool M-dwarfs \citep{2018ExA....45..147C,2019A&A...623A.131K}, showing that it is feasible to detect Ly-$\alpha$ emission because of the strong opacity of the line and the large extent of the Earth’s exosphere. Earth-like planets orbiting at close distances of M dwarf stars are expected to be hit by strong stellar winds, with values that could be comparable to those of young solar-like stars. The ionization state of the exosphere of such these planets depends on the incident photoionizing radiation of the star and the density of the stellar winds. If the radiation and stellar winds of such these stars are strong enough to ionize a large fraction of the present HI exospheric particles, the stellar wind will remove the complete exosphere, as shown in our simulation in the case of young stars. Fainter winds could lead to the formation of large bow shocks at several planetary radii, even though the stellar radiation is strong enough to ionized the upper atmosphere of the planet, regarding to the result of our simulations for more evolved stars.

To summarize, we have carried out numerical MHD simulations of the interaction between the stellar wind coming from solar-like stars of ages 0.1, 1.0, 3.0 and 5.0 Gyr, and an unmagnetized Earth-like planet provided with a H-rich exosphere, considering different initial exospheric densities. From the results of these simulations, we find that the ionized component of the hydrogen planetary exospheres is quickly swept for all considered stellar ages, being these times faster for younger stars. The contribution of the initial exospheric hydrogen density is only relevant for more evolved stars (ages of 3.0 and 5.0 Gyr), where large bow shocks are formed in front of the planet. Hence, we derive that extended exospheres cannot be maintained at early stages.

As a final remark, we would like to note that our simulations are run for a planet orbiting at a fixed distance of 1.0 au around a solar-like star. However, some of our results could be compared to other star-planet configurations: a planet orbiting at 1.0 au from a relatively active star could be comparable to a planet orbiting at very close distances around an older star. The planetary orbital distance is a crucial parameter in the stellar wind-planet interactions, and must be considered in order to obtain a distance-bow shock formation relation. These assumptions will be accurately addressed in future works.

\section*{Acknowledgements}

This work has been funded by the Ministry of Economy and Competitiveness of Spain through grants ESP2015-68908-R and ESP2017-87813-R.

\section*{Data Availability}
The data underlying this article are available in the article and in its online supplementary material.











\bsp	
\label{lastpage}
\end{document}